\documentclass[aps,pre,twocolumn,showpacs]{revtex4}

\pdfoutput=1

\usepackage{amsmath,amssymb,gensymb}
\usepackage{bm}
\usepackage{graphicx}

\usepackage{color}

\usepackage{ulem}

\newcommand{\rev}[1]{\textcolor{black}{#1}}

\begin{document}

\title{
Getting drowned in a swirl: deformable bead-spring model microswimmers in external flow fields
}

\author{Niklas K\"uchler}
\email{kuechler@thphy.uni-duesseldorf.de}
\affiliation{
Institut f\"ur Theoretische Physik II: Weiche Materie, Heinrich-Heine-Universit\"at D\"usseldorf, D-40225 D\"usseldorf, Germany
}

\author{Hartmut L\"owen}
\affiliation{
Institut f\"ur Theoretische Physik II: Weiche Materie, Heinrich-Heine-Universit\"at D\"usseldorf, D-40225 D\"usseldorf, Germany
}

\author{Andreas M.\ Menzel}
\email{menzel@thphy.uni-duesseldorf.de}
\affiliation{
Institut f\"ur Theoretische Physik II: Weiche Materie, Heinrich-Heine-Universit\"at D\"usseldorf, D-40225 D\"usseldorf, Germany
}

\date{\today}

\begin{abstract}
Deformability is a central feature of many types of microswimmers, e.g.\ for artificially generated self-propelled droplets. Here, we analyze deformable bead-spring microswimmers in an externally imposed solvent flow field as simple theoretical model systems. We focus on their behavior in a circular swirl flow in two spatial dimensions. Linear (straight) two-bead swimmers are found to circle around the swirl with a slight drift to the outside with increasing activity. In contrast to that, we observe for triangular three-bead or square-like four-bead swimmers a tendency of being drawn into the swirl and finally getting drowned, although a radial inward component is absent in the flow field. During one cycle around the swirl, the self-propulsion direction of an active triangular or square-like swimmer remains almost constant, while their orbits become deformed exhibiting an ``egg-like'' shape. Over time, the swirl flow induces slight net rotations of these swimmer types, which leads to net rotations of the egg-shaped orbits. Interestingly, in certain cases, the orbital rotation changes sense when the swimmer approaches the flow singularity. Our predictions can be verified in real-space experiments on artificial microswimmers.
\end{abstract}



\maketitle


\section{Introduction}\label{intro}

In recent years, individual self-propelled particles and their collective behavior have been in the focus of intensive research, see Refs.~\cite{lauga2009hydrodynamics,ebbens2010pursuit,romanczuk2012active,cates2012diffusive, menzel2015tuned,elgeti2015physics} for recent reviews. This includes the individual dynamics of
particles with complex shape \cite{kummel2013circular,wensink2014controlling,nguyen2014emergent}, as well as cases of self-rotation \cite{teeffelen2008dynamics,fily2012cooperative,tarama2013dynamics,tarama2013dynamicsshear, nagai2013rotational,nguyen2014emergent,tarama2014individual}. Furthermore, the collective behavior of many such interacting particles has been explored \cite{vicsek1995novel,toner1995long,chate2008collective, deseigne2010collective,schaller2010polar,thutupalli2011swarming, menzel2012collective,kaiser2012how,fily2012athermal, wensink2012meso,menzel2013unidirectional,palacci2013living,buttinoni2013dynamical, menzel2013traveling,svensek2013collective,ihle2013invasion,grossmann2014vortex, kaiser2014transport,speck2014effective,menzel2015focusing}. Collections of self-propelled particles in liquid environment exhibit fascinating
and complex nonequilibrium phenomena emerging from self-organization, where hydrodynamic interactions can play a significant role \cite{hernandez2005transport,cisneros2007fluid,saintillan2008instabilities, saintillan2008pattern,pooley2007hydrodynamic, leoni2010dynamics,gotze2010mesoscale, evans2011orientational,fily2012cooperative, desreumaux2012active,molina2013hydrodynamic,saintillan2013active, alarcon2013spontaneous,menzel2014active,li2014hydrodynamic,zottl2014hydrodynamics, hennes2014self,lushi2014fluid}.

An important example of such microswimmers are colloidal Janus particles \cite{paxton2004catalytic,howse2007self,walther2008janus, jiang2010active,volpe2011microswimmers, buttinoni2012active,theurkauff2012dynamic,buttinoni2013dynamical} that propel via a mechanism of self-induced thermo- or diffusiophoresis \cite{golestanian2007designing}. Due to the surface heterogeneity of Janus particles, they can selectively be heated on one side, or only the coverage of one of the two sides can catalyze chemical reactions. In this way, thermal or concentration gradients build up on length scales of the particle diameter, which in total leads to a net self-propulsion \cite{kitahata2011spontaneous,yoshinaga2012drift,schmitt2013swimming}.

While colloidal Janus particles represent a class of \textit{rigid} (i.e.\ form-stable) artificial microswimmers, several examples of microswimmers were identified that actively \textit{deform} to achieve self-propulsion. Many biological microorganisms fall into this category \cite{brennen1977fluid,wada2007model,polin2009chlamydomonas}, as well as several instances of proposed theoretical model microswimmers \cite{najafi2004simple,golestanian2008analytic, avron2005pushmepullyou,pooley2007hydrodynamic,pande2014micro,pande2015forces}.  
Apart from that, microswimmers can be realized in the form of \textit{soft} particles that are already \textit{deformable} in their non-propelling passive state. In the self-propelling state, such deformations can couple to the migration velocity, which influences the single-particle properties \cite{ohta2009deformable,ohta2009deformation,hiraiwa2010dynamics,tarama2011dynamics, shitara2011deformable,hiraiwa2011dynamics,tarama2013oscillatory, tarama2014individual,menzel2015tuned} as well as their collective behavior \cite{ohkuma2010deformable,itino2011collective,itino2012dynamics,menzel2012soft, tarama2014individual,yamanaka2014formation,lober2015collisions,menzel2015tuned}. Experimental realizations of deformable self-propelled particles are, for instance, migrating cells and self-propelled droplets on rigid surfaces \cite{rappel1999self,lee2002chemical,maeda2008ordered}. Moreover, activated droplets may self-propel over liquid surfaces and through bulk fluids \cite{nagai2005mode,chen2009self,takabatake2011spontaneous,kitahata2011spontaneous, thutupalli2011swarming,schmitt2013swimming,nagai2015self}, which implies induced fluid flows and thus classifies them as deformable microswimmers. An interaction with imposed surrounding flow fields becomes important when rheological properties of active systems are considered \cite{hatwalne2004rheology}, when the features of self-propelled microswimmers are exploited in microfluidic devices \cite{darnton2004moving,toonder2008artificial}, or when microorganisms orient themselves in external shear flows \cite{marcos2012bacterial}. 

Recently, the theoretical studies of the influence of deformability \cite{ohta2009deformable,ohta2009deformation,hiraiwa2010dynamics,tarama2011dynamics, shitara2011deformable,hiraiwa2011dynamics,ferrante2013elasticity, ferrante2013collective,tarama2013oscillatory, ohkuma2010deformable,itino2011collective,itino2012dynamics,menzel2012soft, tarama2014individual,yamanaka2014formation} and of external flow fields \cite{hagen2011brownianshear,zottl2012nonlinear,zottl2013periodic} have been combined, when the behavior of deformable self-propelled particles in a linear shear flow \cite{tarama2013dynamicsshear} and in a swirl flow \cite{tarama2014deformable} has been analyzed. Together with a possible self-spinning motion \cite{tarama2012spinning,nagai2013rotational,tarama2013dynamics}, deformability leads to a multitude of new dynamic states implying different types of winding and cycloidal trajectories, periodic and quasi-periodic motions, or chaotic states \cite{tarama2013dynamicsshear}. The corresponding model equations, especially the couplings between deformation and migration velocity \cite{ohta2009deformable,ohta2009deformation,hiraiwa2010dynamics,tarama2011dynamics, shitara2011deformable,hiraiwa2011dynamics,tarama2013oscillatory, ohkuma2010deformable,itino2011collective,itino2012dynamics,menzel2012soft, tarama2014individual,yamanaka2014formation} as well as between deformation and the external flow \cite{tarama2013dynamicsshear,tarama2014deformable}, are introduced using symmetry arguments, leaving the coupling parameters largely undetermined. This renders the resulting model more general but requires tedious calculations to connect the remaining parameters to real experimental systems \cite{yoshinaga2014spontaneous}. More importantly, so far the back-reaction of the deformable particles on the surrounding fluid flow, which becomes important when many such swimmers act together and interact hydrodynamically, has not been included yet. One could also start directly with a more resolved model that explicitly contains the flow field around each finitely sized model swimmer for this purpose. However, when many such microswimmers interact with each other, the situation becomes increasingly complicated, requiring complex particle- and flow-resolved simulation approaches \cite{evans2011orientational,alarcon2013spontaneous,zottl2014hydrodynamics}.

To cure the missing links in the coupling between finite extension, deformability, and gradients of external flow fields, yet still on a readily accessible level of low complexity, we propose simple minimal bead-spring model microswimmers. It is straightforward to formulate this model, and it can be realized in corresponding experiments. In this case, the parameters of the model are directly related to accessible experimental system parameters. Thermal fluctuations and hydrodynamic interactions within a collection of microswimmers can consistently be included, if necessary. 

\rev{
Here, we focus on the effect that finite size and deformability have on the motion in external flow fields. Due to their finite extension, our swimmers can sense spatial variations in the fluid flow. 
As we demonstrate, this for example leads to overall rotations in swimmer orientations, although we impose a locally irrotational swirl flow. Variations of our simplified approach can serve to effectively and economically describe basic properties of microorganisms in external flows. In fact, aspects of the behavior predicted below have just been found for the motion of bacteria in imposed swirl flows \cite{sokolov2015rapid}. Even the properties of whole colonies of connected microorganisms, such as Volvox colonies \cite{drescher2009dancing,drescher2010direct}, may be characterized accordingly. Corresponding synthetic experimental realizations of our model may serve as experimental model systems. Deformability is contained by construction in our model, but can be reduced by increasing the spring constant. 
}

We introduce our simple deformable bead-spring model microswimmers in section \ref{model}. After that, the simplest realizations of this model in the form of a two-bead, three-bead, and four-bead microswimmer are analyzed in sections \ref{twobead}, \ref{threebead}, and \ref{fourbead}, respectively, in the presence of an externally imposed swirl flow. 
We choose this flow field because it can easily be realized experimentally and leads to interesting aspects of the single-swimmer behavior. 
In section \ref{flucthydro} we discuss possible extensions of our model. 
Finally, we summarize our results and conclude in section \ref{conclusions}.

\section{Deformable model microswimmers}\label{model}

First, we specify our simplified deformable model microswimmers. In our minimal approach, the swimmer body is discretized into an arrangement of $M$ identical spherical beads labeled by an index $i$. Each bead is exposed to frictional forces when it moves relatively to the surrounding fluid with a friction coefficient $\zeta=6\pi\eta R_h$; here $\eta$ is the fluid viscosity and $R_h$ the hydrodynamic radius of the beads. The spherical beads are linked to each other by harmonic springs of spring constant $k$. These springs have finite extension in the undeformed state, which implies a nonzero extension of the deformable microswimmer in its motionless ground state. In the following, we concentrate on microswimmers composed of $M=1$, $2$, $3$, and $4$ beads, referred to as $M$-bead swimmers ($M$-BS). 
\begin{figure}
	\includegraphics[width=7.cm]{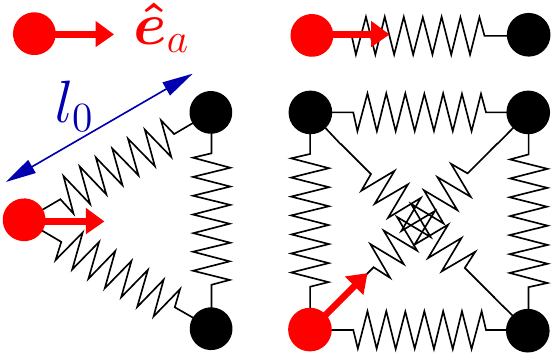}
	\caption{\rev{(Color online)} Minimal deformable bead-spring model microswimmers in their undeformed motionless ground states. The spherical beads are connected via harmonic springs of length $l_0$ in the undeformed state ($\sqrt{2}l_0$ for the diagonal springs in the square). The objects are referred to as $M$-bead swimmers ($M$-BS), where here $M=1,2,3,4$. In the active case, one of the beads (red/brighter) is self-propelled with an active drive $\boldsymbol{a}=a\mathbf{\hat{e}}_a$. For $M\geq2$, $\mathbf{\hat{e}}_a$ points towards the center of mass of the other (passive) beads. }
	\label{fig:swimmer_model}
\end{figure}
Fig.~\ref{fig:swimmer_model} illustrates the considered regular swimmer geometries. In the undeformed state, the $2$-BS represents a linear object, the $3$-BS an equilateral triangle, and the $4$-BS a square. All springs feature a length $l_0$ in the undeformed state, except for the additional diagonal springs of undeformed lengths $\sqrt{2}l_0$ that counteract shear deformations of the square. 

In the active case, one of the beads, labeled by an index $m$, features an active drive $\boldsymbol{a}=a\mathbf{\hat{e}}_a$. For $M\geq2$, we let the unit vector $\mathbf{\hat{e}}_a$ point towards the center of mass of all other (passive) beads. Thus, the orientation of the active drive $\boldsymbol{a}$ is set within the body frame of each deformable swimmer and is not imposed from outside. Each swimmer is therefore an autonomous self-driven entity. $|a|$ sets the strength of the active drive, while the unit vector $\mathbf{\hat{e}}_a$ and the sign of $a$ determine the direction and orientation of the active drive. For $a>0$, the active drive is oriented ``inward'', therefore the active bead ``pushes'' the other beads. In contrast to that, for $a<0$, the active drive points ``outward'', thus the active bead ``pulls'' on the remainder of the swimmer.  

\rev{
At this point, we should insert a comment to make clear the difference between our approach on the one hand and previous models of linked-sphere microswimmers on the other hand \cite{najafi2004simple,golestanian2008analytic, avron2005pushmepullyou,pooley2007hydrodynamic,pande2014micro,pande2015forces}. In those previous models, passive spheres were considered as building blocks. Activity was introduced via the links between them, i.e.\ actively contracting and expanding joints or springs connecting the spheres. Only due to the work of those active links did the assembled swimmer objects self-propel, for instance because of overall non-reciprocal deformation cycles. In contrast to that, we here consider passive elastic springs as links. However, one of the beads already by itself is active and self-propelling. In reality, it may be given, for example, by a rigid self-propelling spherical Janus particle. Such Janus beads self-propel without the aid of any mechanically active joints, but via phoretic self-induced surrounding temperature or concentration gradients \cite{paxton2004catalytic,howse2007self,walther2008janus, jiang2010active,volpe2011microswimmers, buttinoni2012active,theurkauff2012dynamic,buttinoni2013dynamical, golestanian2007designing,kitahata2011spontaneous, yoshinaga2012drift,schmitt2013swimming}. In the following, we only consider one active bead per swimmer entity. Naturally, also several active beads connected by elastic springs can be considered \cite{kaiser2015does,chen2015transport,babel2015dynamics}.
}

Typically, the dynamics of microswimmers is restricted to the regime of low Reynolds numbers \cite{purcell1977life}. Therefore we consider an overdamped type of dynamics.
\rev{Hydrodynamic interactions between the individual beads composing one swimmer body are neglected in the present approach, where we concentrate on the effect of deformability and finite size of the swimmers. This is in accord with a lowest-order expansion in the ratio between the bead size and the typical distance between the beads; see also section \ref{flucthydro} for further comments on this point. It has been demonstrated in detail that the behavior of an individual, not hydrodynamically interacting microswimmer can be described using an \textit{effective} active drive ($\boldsymbol{a}=a\mathbf{\hat{e}}_a$ in the present context) \cite{kummel2014kummel,hagen2015can}. This concept does \textit{not} contradict the notion of an overall force-free microswimmer. An illustrative way to understand this perception is the following. A microswimmer is trapped in a confining external potential. It will not stay in the potential minimum. Instead, it moves against and climbs up the potential walls \cite{nash2010run,hennes2014self,menzel2015focusing,menzel2015dynamical}, at least as long as it does not reorient. Being able to work against the walls requires an effective active drive of the swimmer. The swimmer comes to rest when this active drive is balanced by the counteracting potential force. In this ``stuck'' situation, the active drive becomes visible: it is now transmitted to the fluid, which is set into motion as if a net force were acting on it. An effective hydrodynamic fluid pump can be realized in this way \cite{nash2010run,hennes2014self,menzel2015dynamical}.
We apply this approach to the one active bead on our swimmer body.}

Combining all these ingredients, we find the following equation of motion for the $i$-th bead: 
\begin{equation}
\label{eq_motion_unscaled}
 \zeta \left(\frac{\text{d}{\bf r}_i}{\text{d}t} - {\mathbf{u}({\bf r}_i)}\right) = a\mathbf{\hat{e}}_a\,\delta_{im} + \sum_{\substack{ j=1 \\ j\neq i} }^{M} {\bf f}_{ij},\quad i = 1,\text{...},M.
\end{equation}
On the left-hand side, this equation lists the linear viscous friction of the $i$-th bead with its fluid environment. $\mathbf{r}_i$ denotes the position of the $i$-th particle and $\mathbf{u}(\mathbf{r})$ the flow field of the surrounding fluid. Therefore, the brackets contain the relative velocity of the $i$-th bead with respect to the surrounding fluid. On the right-hand side, the first term includes the active drive, with $\delta_{im}$ the Kronecker delta. The second term contains the pairwise interaction forces between different beads. In our case, these result from the harmonic springs and are of the form 
\begin{equation}
 {\bf f}_{ij} = k \left(\|{\bf r} _{ij}\| - l_{ij}\right) {\mathbf{\hat{r}}} _{ij}, \quad i,j = 1,...,M.
\end{equation}
Here, $l_{ij}$ is the length of the undeformed spring connecting beads $i$ and $j$, $\mathbf{r}_{ij}=\mathbf{r}_j-\mathbf{r}_i$, and $\mathbf{\hat{r}} _{ij}=\mathbf{r}_{ij}/\|\mathbf{r}_{ij}\|$. In this work, we will not consider situations that would make it necessary to introduce steric interactions between the beads. Furthermore, we neglect possible couplings between rotations of the beads and spring deformations \cite{pessot2015towards}. 

In this investigation, we study the dynamic behavior of such model microswimmers in two spatial dimensions. Experimentally, this could be realized by tracking the motion of such swimmers on the surface of a liquid. Another option would be to confine the motion of microswimmers to the interface between two immiscible fluids. 

Within the accessible two-dimensional area, we consider a prescribed flow field $\mathbf{u}(\mathbf{r})$ given by 
\begin{equation}\label{eq:flow_field}
 {\bf{u}}({\bf r}) = \frac{\lambda}{r} {\hat {\bf e}}_\phi ({\bf r})
\end{equation}
in polar coordinates. It describes a swirl flow around the origin, with $\nabla\cdot\mathbf{u}=0$ for ${\bf{r}} \neq \bf 0$, corresponding to an incompressible fluid flow. The sign of $\lambda$ sets the rotational orientation of the swirl, while $|\lambda|$ characterizes its strength. $r=\|\mathbf{r}\|$ denotes the distance from the swirl center, $\phi$ the azimuthal angle, and ${\hat {\bf e}}_\phi ({\bf r})$ the unit vector in the azimuthal direction. For ${\bf{r}} \neq \bf 0$ we have ${\bf{\nabla}} \times {\bf u} = \bf 0$ so that the flow field can locally be derived from a potential field $U(\mathbf{r})$ as $\mathbf{u}(\mathbf{r})=-\nabla U(\mathbf{r})$, where $U(\mathbf{r})=\lambda\arctan(x/y)$ in Cartesian coordinates.  

The study of this particular flow field offers two major advantages. On the one hand, and in contrast to the often analyzed linear shear profile \cite{hagen2011brownianshear,tarama2013dynamicsshear}, a swirl flow can easily be realized experimentally. In the simplest case, a swirl can be induced in a cylindrical cavity of sufficient diameter and height by a rotating magnetic stir bar at the bottom of the vessel. On the other hand, a previous theoretical investigation of a different deformable swimmer model has already demonstrated rich dynamics in this situation \cite{tarama2014deformable}. In that previous study, extensional contributions of the surrounding fluid flow could induce deformations of the swimmer. However, the microswimmer itself was approximated as a point-like object. In the present model, we consider a finite extension of our swimmers within the gradient of the externally imposed flow field. As we will demonstrate, this leads to qualitatively new types of behavior. 

Finally, we switch to dimensionless units by rescaling $\mathbf{r}'=\mathbf{r}/l_0$, $\mathbf{r}_i'=\mathbf{r}_i/l_0$, $l_{ij}'=l_{ij}/l_0$, $t'=t\lambda/l_0^2$, ${a}'={a}l_0/\zeta\lambda$, and $k'=kl_0^2/\zeta\lambda$. 
Thus $l_{ij}'=1$, except for the diagonal springs in the $4$-BS, where $l_{ij}'=\sqrt{2}$. Omitting the primes in the following, we are left with the equations of motion 
\begin{eqnarray}
 \frac{\text{d}{\bf r}_i}{\text{d}t} &=& {\mathbf{u}({\bf r}_i)}+ a\mathbf{\hat{e}}_a\,\delta_{im} + k \sum_{\substack{ j=1 \\ j\neq i} }^{M} \left(\|{\bf r} _{ij}\| - l_{ij}\right) {\mathbf{\hat{r}}} _{ij},\nonumber\\
{\bf{u}}({\bf r}) &=& \frac{1}{r} {\hat {\bf e}}_\phi ({\bf r}),\qquad i = 1,\text{...},M.
\label{eq_motion}
\end{eqnarray}
The only two remaining system parameters are the rescaled strength of self-propulsion $a$ and the rescaled spring constant $k$. 

In the following, we present our results from numerical investigations in two spatial dimensions. For this purpose, Eq.~(\ref{eq_motion}) was iterated forward in time using a fourth-order Runge-Kutta scheme \cite{press1992numerical}. The time step was reduced until further changes in the trajectories became negligibly small.

\section{Deformable two-bead microswimmer in the swirl}\label{twobead}

As a first explicit example of our bead-spring swimmer model, we start with the $2$-BS illustrated in Fig.~\ref{fig:swimmer_model}. 
\begin{figure}
	\includegraphics[width=7.5cm]{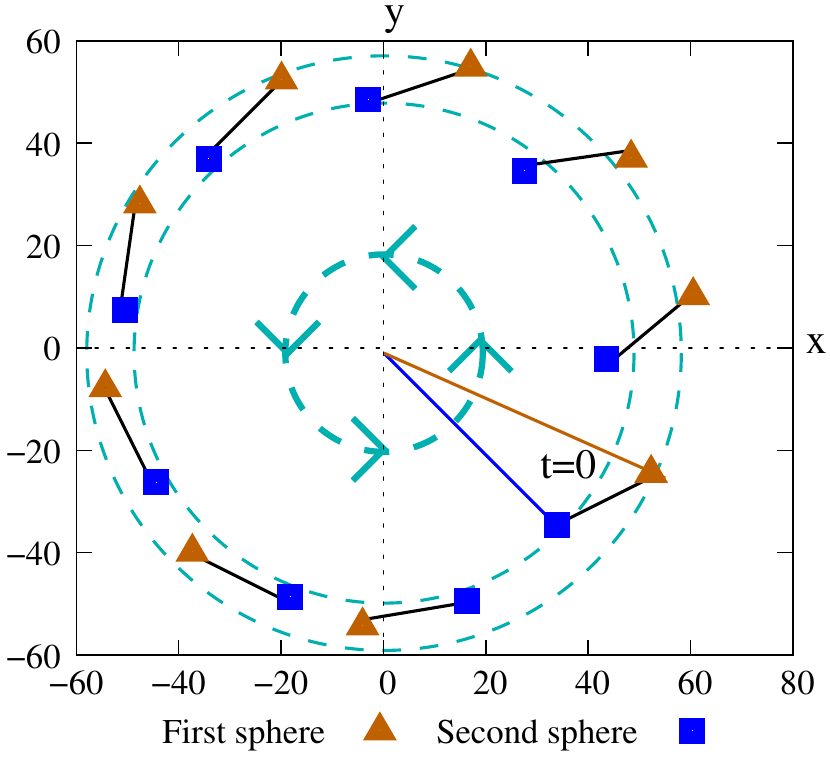}
	\caption{\rev{(Color online)} Tangential alignment process of a passive $2$-BS in the swirl with the circular flow lines. The initial position and orientation is marked by ``$t=0$''. After a short transient, the initially inner bead (square) overtakes the initially outer bead (triangle) and then moves ahead.
	This behavior results from the radial gradient in the flow velocity. 
        Initially, one of the beads (square) is closer to the swirl center than the other one (triangle).
	Therefore, the inner bead (square) is convected faster by the swirl. 
	The swimmer rotates until both beads have the same distance from the swirl center, see the bottommost configuration. For illustration, the extension of the microswimmer is enlarged by a factor of $20$. [Parameters: $a=0$, $k=5$.]}
	\label{fig:passive2BS}
\end{figure}
For $a=0$, we obtain a passive object that is simply convected by the swirl flow. To minimize the deformation energy of the spring connecting the two beads, this passive swimmer aligns with the circular flow lines, see Fig.~\ref{fig:passive2BS}. Due to the finite extension of the swimmer, the geometric role of the spring connecting the two beads becomes that of a secant to the corresponding flow line. In this way, the spring remains undeformed and the passive swimmer performs circular trajectories. The radius of the trajectories is solely determined by the initial conditions and is marginally stable. A similar situation was found for the deformable passive object investigated in Ref.~\cite{tarama2014deformable}. 

\begin{figure}
	\includegraphics[width=7.5cm]{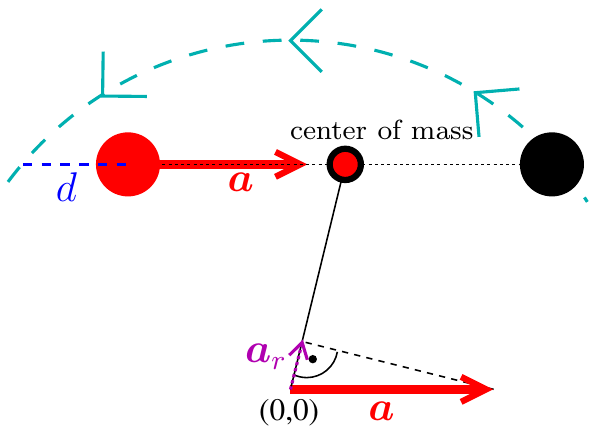}
	\caption{\rev{(Color online)} Schematic illustration of an active $2$-BS in the swirl flow. The active drive $\boldsymbol{a}$ always orients against the circular flow field as further discussed in the main text. At a certain time, for a flow line passing through one of the beads, we consider the secant running through the swimmer axis. Due to the active drive, the swimmer is shifted along this secant such that the flow line misses the other bead by a finite distance $d\neq0$. Due to this shift, when decomposing $\boldsymbol{a}$ as it acts on the swimmer center, a radial component $\boldsymbol{a}_r$ results. Here, we depict the case of $a>0$. 
The principle is the same for $a<0$, where the active bead is located on the opposite end and $\boldsymbol{a}$ points away from the swimmer center. For illustration, we exaggerated the curvature of the flow line with respect to the swimmer extension. }
	\label{fig:active_2BS}
\end{figure}

For the active $2$-BS of low enough $|a|$, we still find nearly circular trajectories. Again, the swimmer axis approximately forms a secant to the local fluid flow lines. Moreover, we observe that the active drive $\boldsymbol{a}=a\mathbf{\hat{e}}_a$ always aligns oppositely to the circular flow lines of the swirl. See Fig.~\ref{fig:active_2BS} for an illustration in the case of $a>0$.
Considering the swimmer as it is convected forward along the swirl flow, it becomes clear why the active bead in Fig.~\ref{fig:active_2BS} for $a>0$ sits at the front and not at the rear. (The words ``front'' and ``rear'' mark the positions with respect to the convection direction given by the swirl). A situation with the active bead at the rear is not stable. First, due to the active drive, the active bead becomes faster and tends to overtake the passive bead. Second, when it does so, due to the secant geometry it pushes towards regions of higher swirl flow velocity, which further adds to the overtaking procedure. In the end, we find the active bead at the front for $a>0$. 

The opposite case of $a<0$ is likewise very intuitive. Then the active drive points away from the swimmer center. In this situation, again considering the forward-convection by the swirl, the active bead sits at the rear. It slows the swimmer convection. Having the brake at the rear is a stable configuration. 

\begin{figure}
	\includegraphics[width=7.cm]{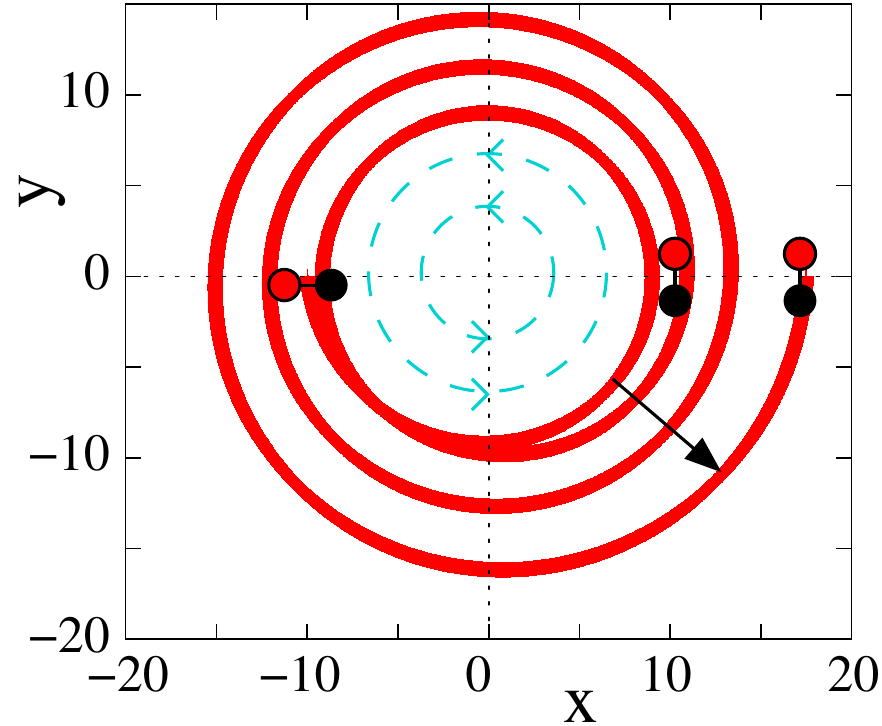}
	\caption{\rev{(Color online) Example trajectory for an active $2$-BS in the swirl (here for $a>0$, but similarly for $a<0$). After an initial reorientation process, the swimmer aligns with the flow field as illustrated in detail in Fig.~\ref{fig:active_2BS}. Swimmer orientations are indicated by the insets, where the passive bead is marked in black. The active drive then leads to an outward drift as becomes evident from the spiral-like trajectory. As a result, the active $2$-BS does not get drowned in the swirl center. [Parameters: $a=0.019$, $k=0.5$.]}}
	\label{fig:trajectory_active_2BS}
\end{figure}

In both cases, $a>0$ and $a<0$, we observe that over time the swimmer is slowly drifting radially outwards. \rev{A corresponding trajectory is depicted in Fig.~\ref{fig:trajectory_active_2BS}.} 
\rev{This phenomenon} is easily understood when we recall that the active drive is always oriented against the flow lines of the swirl, see Fig.~\ref{fig:active_2BS}. The active drive $\boldsymbol{a}$ drags the swimmer along the secant passing through the swimmer axis. Thus, for $a\neq0$, the two beads are not located on the same circular flow line. When viewed from the swimmer center, this shift leads to an effective radial component $\boldsymbol{a}_r$ resulting from the active drive. It drags the swimmer radially outwards. \rev{Related effects are reported in recent experimental studies \cite{sokolov2015rapid}.} As a consequence, the active $2$-BS is safe from getting drowned in the swirl.

\section{Deformable three-bead microswimmer in the swirl}\label{threebead}

Next, we turn to the $3$-BS. It features a much richer dynamics than the simple $2$-BS, thus we are significantly more explicit here. For not too high strengths of the active drive, the $3$-BS shows a characteristic opposite behavior when compared to the $2$-BS in that it gets drawn into the swirl. This property is already observed for the passive $3$-BS and is traced back to its two-dimensional spatial extension. Thus, a low-powered $3$-BS is getting drowned in the swirl. Only for strong enough active drive, the $3$-BS can escape the swirl. 

\subsection{Passive three-bead microswimmer}\label{sec:passive-3BS}

We start by investigating the passive case $a=0$. Since by construction of our model we explicitly take into account the finite spatial extension of the swimmer, qualitatively new effects arise when compared to previous descriptions \cite{tarama2014deformable}. In particular, the swimmer, while being convected along approximately circular trajectories around the swirl, is slowly but persistently drawn into the swirl center. 

\begin{figure}
	\includegraphics[width = 8.cm]{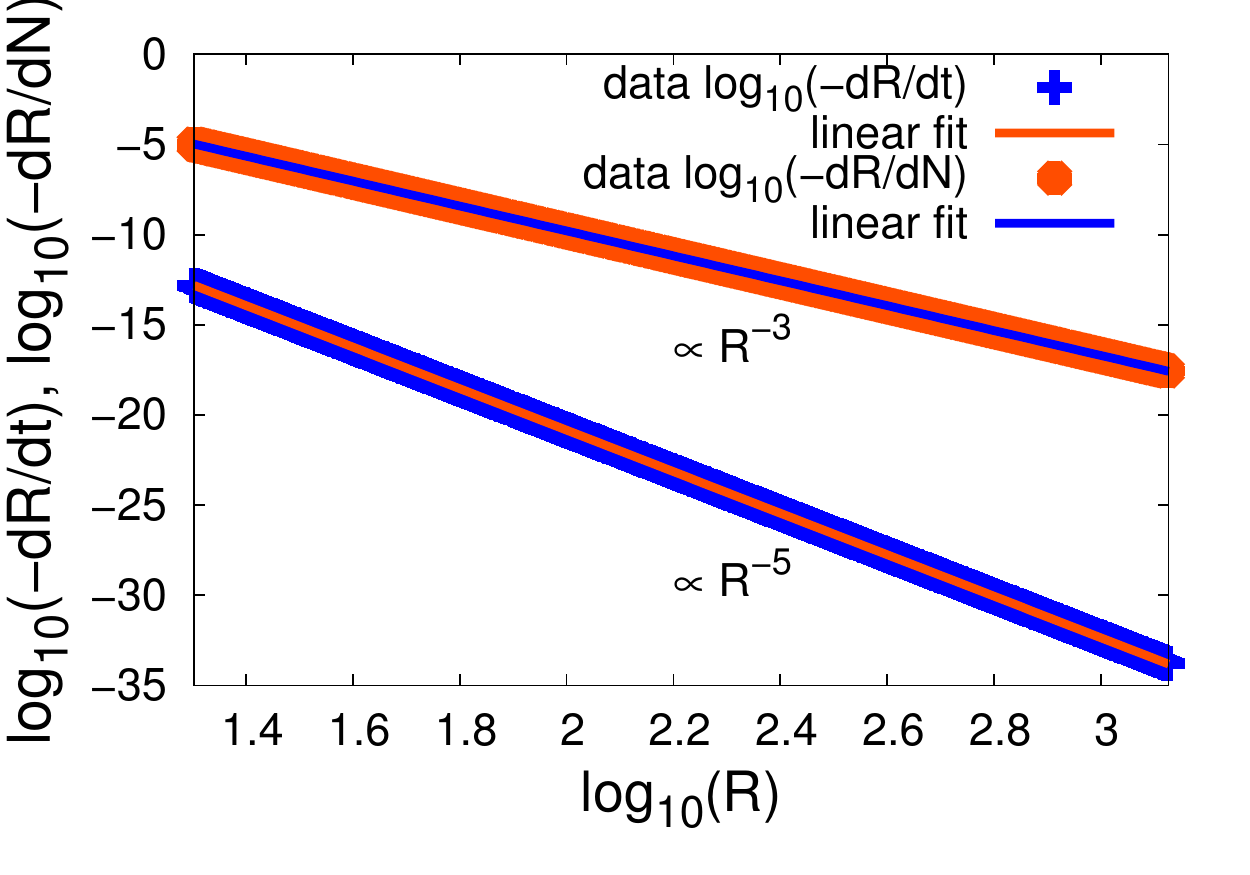}
	\caption{\rev{(Color online)} Undertow velocity toward the swirl center as a function of the distance $R$ of the passive $3$-BS from the swirl center. 
We plot the undertow velocity with respect to the passed time $t$ 
and with respect to the number $N$ of circles around the swirl. 
The plot suggests power laws $\mathrm{d}R/\mathrm{d}t\propto R^{-5}$ and $\mathrm{d}R/\mathrm{d}N\propto R^{-3}$. The lines represent fits of the data points according to these power laws. [Parameters: $a=0$, $k=0.05$.] 
    }
	\label{fig:sog_geschwind}
\end{figure}

Denoting the center-of-mass position by $\mathbf{R}$ and its distance from the swirl center by $R=\|\mathbf{R}\|$, we obtain from a fit to our numerical results the algebraic relation
\begin{equation}
\frac{\mathrm{d}R}{\mathrm{d}t}=-bR^{m_1} 
\end{equation}
for the decrease of the distance from the swirl center. The corresponding numerical fitting parameters 
\begin{equation}
\ln(b) \approx 4.020, \qquad m_1\approx-5.000
\end{equation}
were determined to three-digit precision. We obtained the same value for $m_1$ independently of the chosen spring constant $k$. 
Likewise, this result can be quantified using as a variable the number of cycles around the swirl center $N$ instead of time $t$. Then we obtain
\begin{equation}
\frac{\mathrm{d}R}{\mathrm{d}N}=-cR^{m_2} 
\end{equation}
with 
\begin{equation}
\ln(c) \approx 2.183, \qquad m_2\approx-3.000.
\end{equation}
Assuming virtually spherical trajectories around the swirl center, we analytically obtain the relations $m_1=m_2-2$ and $c=2\pi b$ in agreement with the above numerical values. 
We show examples for the decay rates $\mathrm{d}R/\mathrm{d}t$ and $\mathrm{d}R/\mathrm{d}N$ in Fig.~\ref{fig:sog_geschwind}, where the above relations were found to hold at least across two orders of magnitude in the distance $R$. 
In the present framework, our description remains meaningful only for distances $R$ larger than the extension of the microswimmer. 

\begin{figure}
	\includegraphics[width=8.cm]{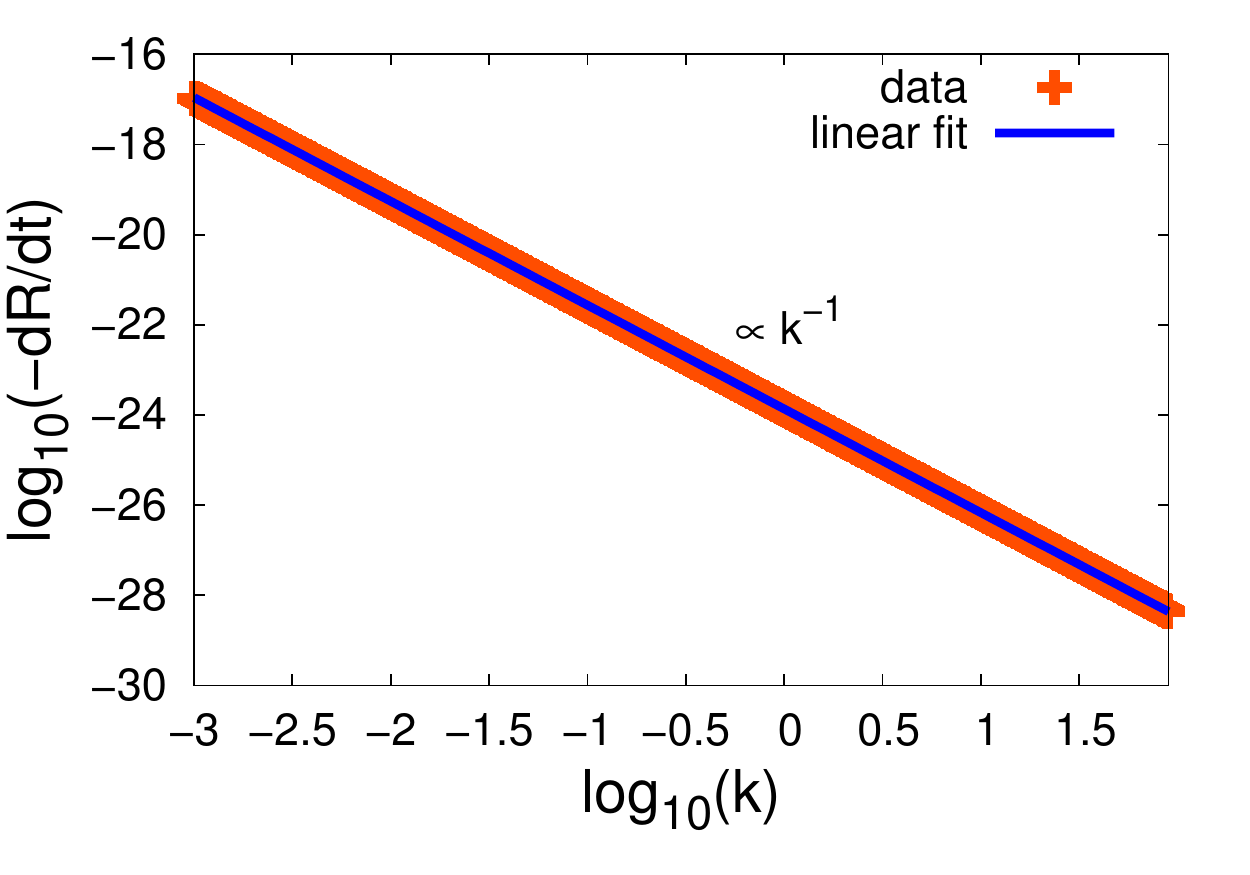}
	\caption{\rev{(Color online)} Undertow velocity ${\text{d} R}/{\text{d}t}$ of the passive $3$-BS as a function of the spring constant $k$. 
The plot suggests a power law ${\text{d} R}/{\text{d}t}\propto k^{-1}$. The line represents a fit of the data points according to this power law.  [Parameters: $a=0$, $R(t=0)= 50$.] 
        }
	\label{fig:sog_federkonstante}
\end{figure}

The reason for the observed undertow is the finite extension of the swimmer together with its deformability. Along its extension, the swimmer experiences the gradient of the imposed swirl flow. Inner beads are convected quicker on narrower paths of higher curvature around the swirl center than outer beads. For geometric reasons, this difference in speeds of convection together with the curved trajectories leads to an approach of the center of mass towards the swirl center. That simple picture best applies, if the beads are convected rather independently of each other, i.e.\ for small spring constants $k$. In fact, we observe a decreasing undertow for increasing spring constant according to
\begin{equation}
\frac{\mathrm{d}R}{\mathrm{d}t}\propto k^{-1}  
\end{equation}
over several orders of magnitude, see Fig.~\ref{fig:sog_federkonstante}.

We obtain further evidence for the importance of the gradient in the flow field when we analyze the deformation cycles of the springs. During each cycle around the swirl center, each spring is periodically contracted and extended twice, in agreement with the varying orientation of the spring relatively to the swirl center. As expected, the amplitude of deformation is proportional to the gradient of the flow field, which via Eq.~(\ref{eq:flow_field}) implies a proportionality $\propto R^{-2}$.  

Next, we address the orientation of the entire swimming object. 
\begin{figure}
	\includegraphics[width=7.5cm]{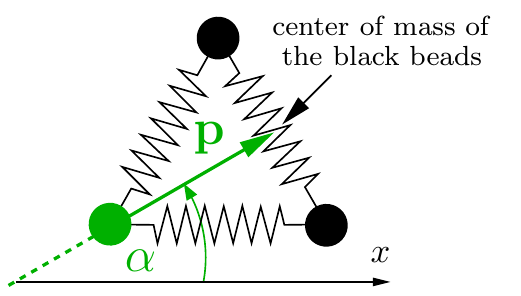}
	\caption{\rev{(Color online)} Definition of the orientational angle $\alpha$ of a $3$-BS. The vector $\bf p$ points from one labeled bead (green/brighter) to the center of mass of the two other beads (black). The same procedure is performed for the two other beads, where the two resulting vectors are rotated by $120\degree$ and $240\degree$, respectively, towards $\bf p$. Here, due to the regular shape, they then all coincide with $\bf p$. We normalize these three vectors and calculate their average. $\alpha$ is defined as the orientational angle of this averaged vector with respect to the $x$-axis of our fixed laboratory frame. } 
	\label{fig:orientierung}
\end{figure}
For this purpose, we introduce an orientational angle $\alpha$ as depicted in Fig.~\ref{fig:orientierung}: first, the vector pointing from one labeled bead to the center of mass of the two other beads is determined and normalized; the analogous vectors are identified starting from the two other beads; these two latter vectors are rotated by $120\degree$ and $240\degree$, respectively, towards the first vector; finally, the average of the three resulting vectors is calculated and its orientational angle $\alpha$ with respect to a fixed laboratory axis is determined. 

\begin{figure}
	\includegraphics[width=8.3cm]{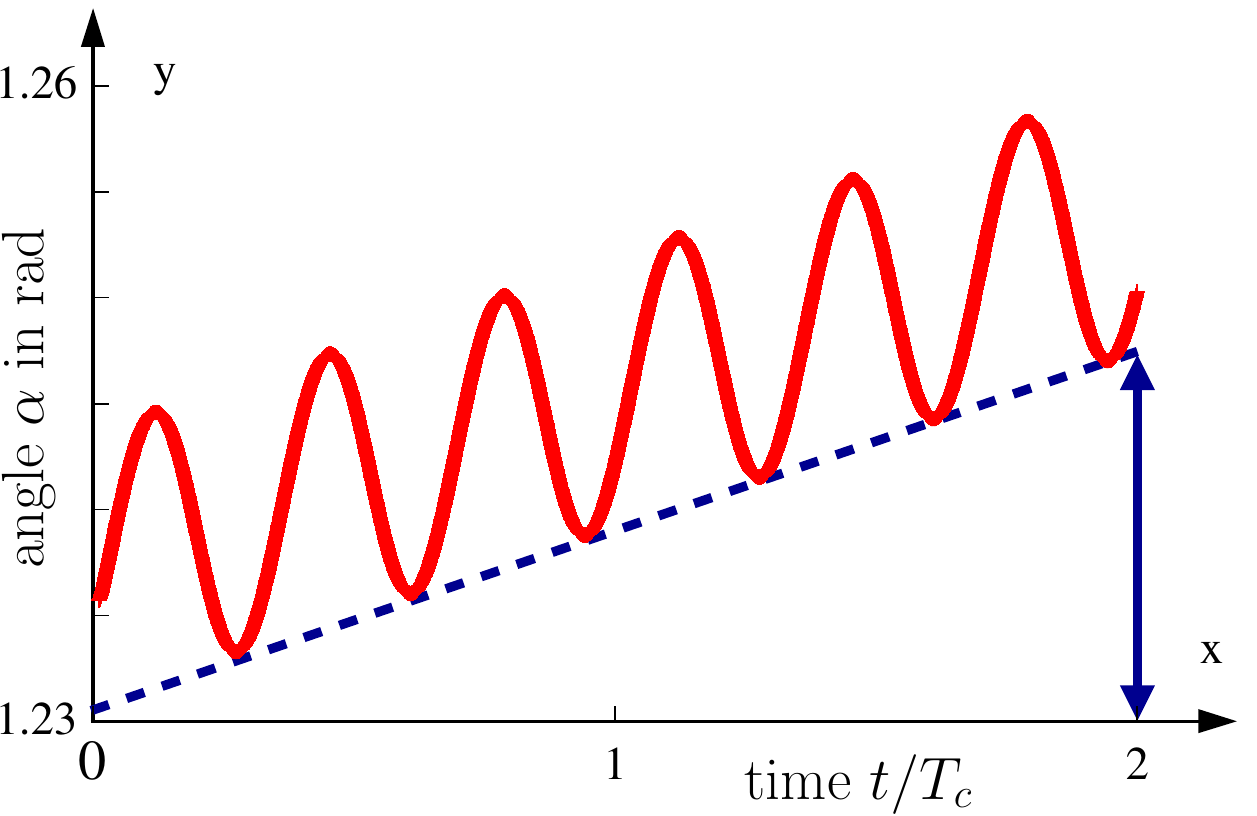}
	\caption{\rev{(Color online)} Angular orientation $\alpha$ of a passive $3$-BS as a function of time $t$. $T_c$ denotes the time necessary for the swimmer to circle around the swirl once. We observe three tiny sinusoidal oscillations in the angular orientation during each cycle. With proceeding time, a net rotational drift takes place, marked by the dashed line.  [Parameters: 
        $a=0$, $k = 0.03$, $R(t=0)=30$, $T_c \approx 5870$.]
    }
	\label{fig:angle_3PS}
\end{figure}

\begin{figure}
	\includegraphics[width=7.5cm]{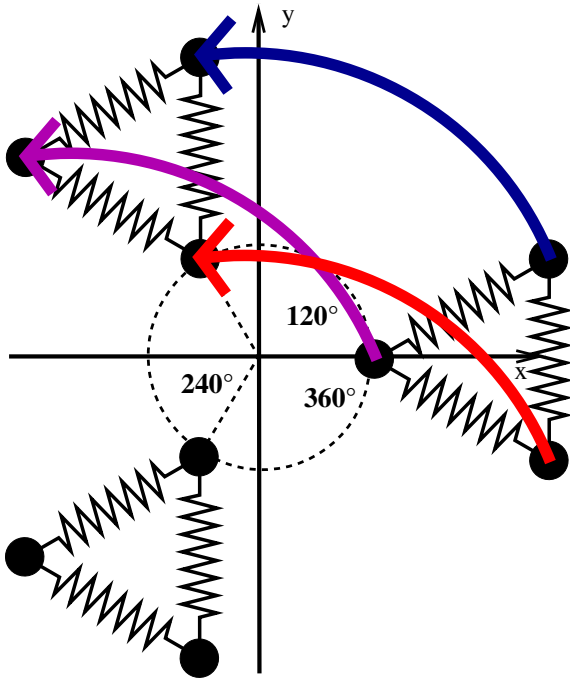}
	\caption{\rev{(Color online)} Three-fold configurational symmetry of a passive $3$-BS while being convected around the swirl flow. To first approximation, the microswimmer is convected along a circular orbit and its orientation remains fixed in space as indicated by the arrows (in accord with the tiny magnitudes of rotation in Fig.~\ref{fig:angle_3PS}). Due to its three-fold rotational symmetry, the passive $3$-BS passes through three symmetry points along one circular orbit (shifted to each other by $120\degree$), from where the physics is the same in each case. For illustration, the size of the microswimmer is significantly exaggerated when compared to the distance from the swirl center.}
\label{fig:dreieck_symmetrie}
\end{figure}

Fig.~\ref{fig:angle_3PS} shows a typical example for the orientational angle $\alpha$ of the passive $3$-BS while it is circling around the swirl center. 
As can be inferred from the scale of the ordinate, orientational changes are tiny, at least in the parameter ranges that we investigated. To first approximation, the orientation of the swimmer remains fixed with respect to the laboratory frame over the depicted time interval. Yet, there is a systematic oscillation in the orientation, with approximately three oscillations during one cycle around the swirl center. This number agrees with the three-fold rotational symmetry of the dynamics while the passive $3$-BS is convected around the swirl, see Fig.~\ref{fig:dreieck_symmetrie}. 

Furthermore, we observe a finite gradual net rotation of the whole swimmer in Fig.~\ref{fig:angle_3PS} caused by its finite extension.
%
From our analysis of the situation, we conclude that this net rotation represents a remnant of the passive $2$-BS rotation dynamics. While the $2$-BS is convected around the swirl center, its body axis as seen from the laboratory frame performs a net rotation. See Fig.~\ref{fig:passive2BS} after the initial transient has decayed. 
Each time the $3$-BS adopts an 
orientation with one particle to the outside and two particles to the inside with equal distance from the swirl center, 
for the two inner particles the situation of the $2$-BS is restored. 
The third outer particle hinders the rotation. 
Yet, the drag on the inner particles is a little higher due to the radial gradient in the flow velocity. In the resulting competition of two against one with higher drag on the inner particles, a small net rotation apparently survives. 
The tiny net rotation of the swimmer is \rev{counterclockwise}, in agreement with the rotational sense of the swirl.

\subsection{Active three-bead microswimmer captured by the swirl}
\label{drawn_active_3BS}

\begin{figure}
		\includegraphics[width=8.cm]{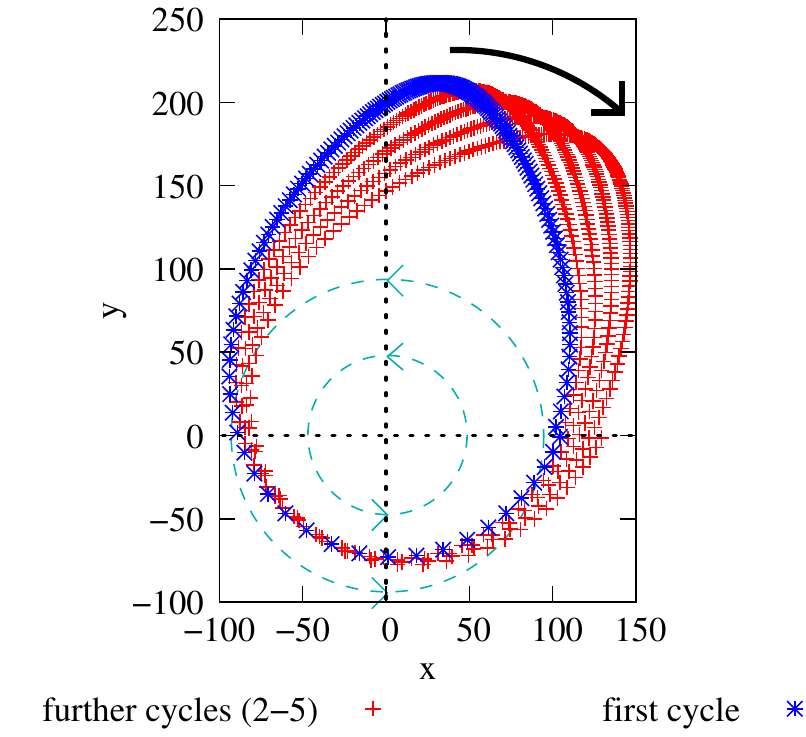}
		\caption{\rev{(Color online)} Center-of-mass trajectory of an active $3$-BS in the swirl when viewed from the laboratory frame. Due to the active drive, the circular orbit of the passive $3$-BS is deformed to an egg-like shape. Over time, due to a net orientational drift of the swimmer, this egg rotates. As depicted, for $a>0$, we could observe a rotational sense of this egg opposite to the rotational sense of the swirl flow. [Parameters: 
                $a = 0.01$, $k=0.1$.]} 
		\label{fig:rotating_egg}
	\end{figure}

At low enough strengths $|a|$ of the active drive, an active $3$-BS placed sufficiently close to the swirl center such that $|a|\ll\|\mathbf{u}\|$ remains captured by the swirl. As observed for the passive $3$-BS, the active $3$-BS then gets drawn towards the swirl center. Except for a little net orientational drift and tiny orientational oscillations (see Fig.~\ref{fig:angle_3PS} for the passive case) the orientation of the active $3$-BS remains approximately constant during one cycle. Thus, during one cycle around the swirl, the active drive always works into the same direction when viewed from the laboratory frame. This leads to a distortion of the orbit from an approximate circle to an ``egg-like'' shape, see Fig.~\ref{fig:rotating_egg}.

\begin{figure}
	\includegraphics[width=7.5cm]{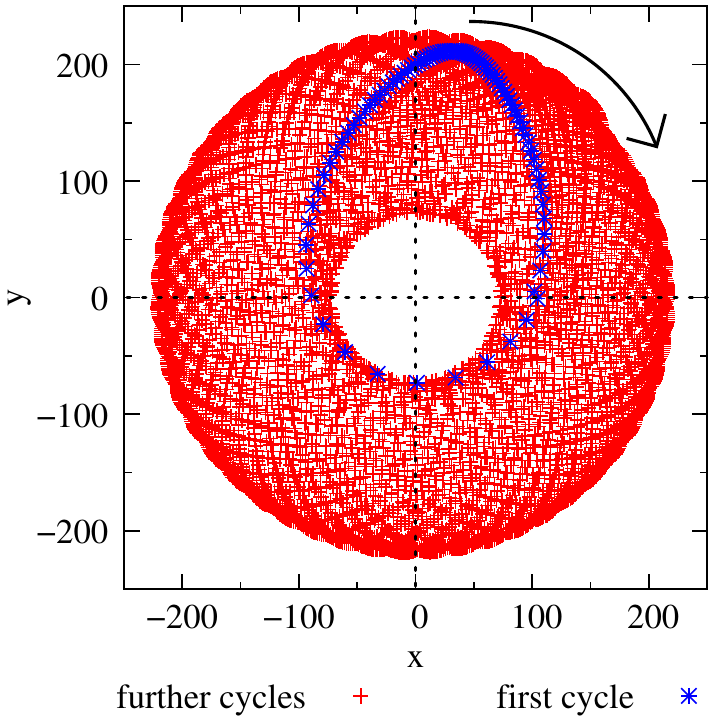}
	\caption{\rev{(Color online)} Same situation as in Fig.~\ref{fig:rotating_egg}, observed over a longer time interval. A rosette-like structure of the trajectory is found. [Parameters: 
                $a = 0.01$, $k=0.1$.]}
	\label{fig:rosette}
	\end{figure}

Already in the passive case, we have observed a small net orientational drift over time. For the active $3$-BS, an orientational drift leads to a rotation of the egg-shaped orbit around the swirl center, see Fig.~\ref{fig:rotating_egg}. Over time, a rosette-like structure arises due to this rotation as shown in Fig.~\ref{fig:rosette}.
%

We remark, however, that the rotational sense of the egg-shaped orbit in Figs.~\ref{fig:rotating_egg} and \ref{fig:rosette} is opposite to the rotational sense of the swirl flow. This is in contrast to what we have found for the passive $3$-BS. Moreover, we only observed this behavior for $a>0$, i.e.\ when the active bead pushed towards the swimmer center. 

\begin{figure}
	\includegraphics[width=8.3cm]{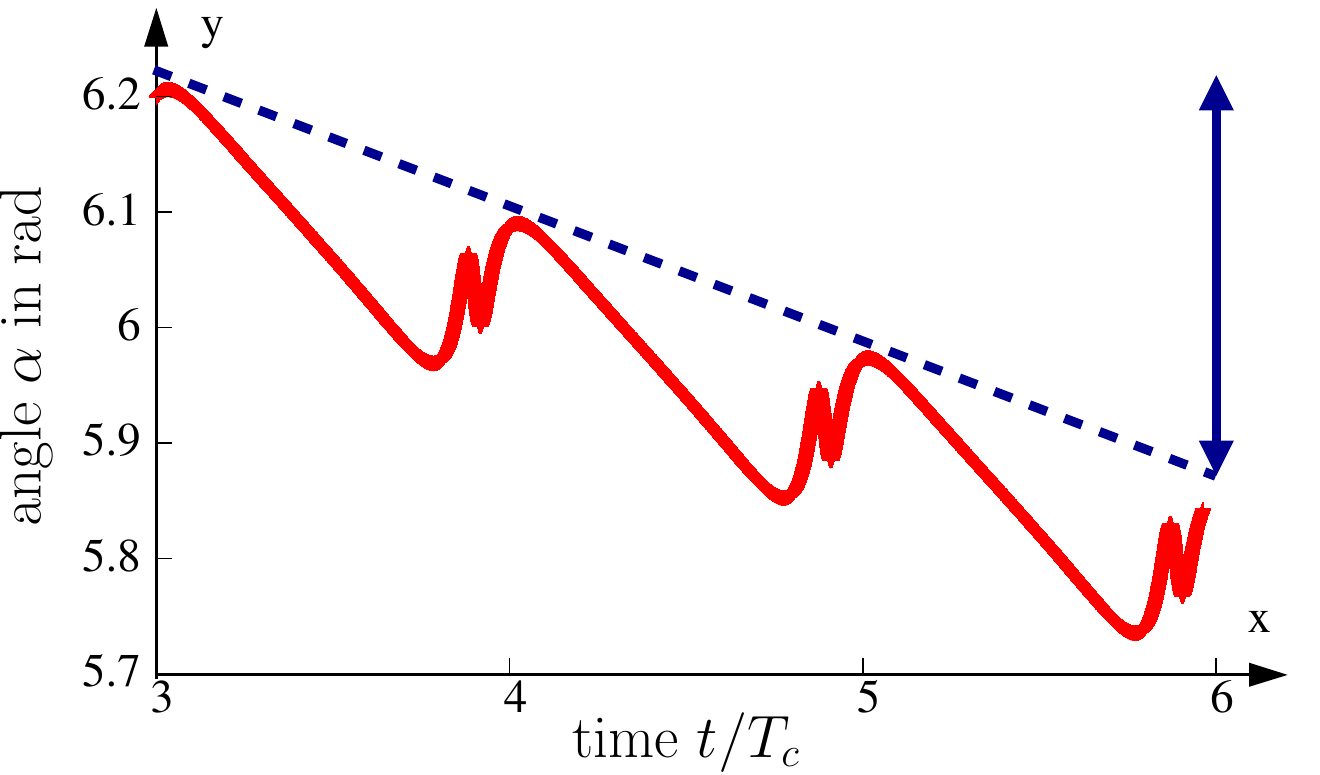}
	\caption{\rev{(Color online)} 
        Angular orientation $\alpha$ of the active $3$-BS in Figs.~\ref{fig:rotating_egg} and \ref{fig:rosette} as a function of time $t$. $T_c$ again denotes the time necessary for the swimmer to circle around the swirl once. During each cycle around the swirl, an extreme event of rotation opposite to the rotational sense of the swirl flow occurs. This event determines the rotational sense of the egg-shaped orbits in Figs.~\ref{fig:rotating_egg} and \ref{fig:rosette}. [Parameters: 
        $a = 0.01$, $k=0.1$, $R(t=0)=100$, $T_c \approx 250000$.]
        }
	\label{fig:angle_3PS_active}
\end{figure}

Fig.~\ref{fig:angle_3PS_active} resolves the orientational behavior with higher resolution. 
We find oscillations together with a rotational drift similar to the passive $3$-BS in Fig.~\ref{fig:angle_3PS}. Yet, during each cycle around the swirl, a pronounced rotation with a sense opposite to the rotational sense of the swirl flow arises. This pronounced event determines the overall rotational appearance. 

\begin{figure}
	\includegraphics[width=7.3cm]{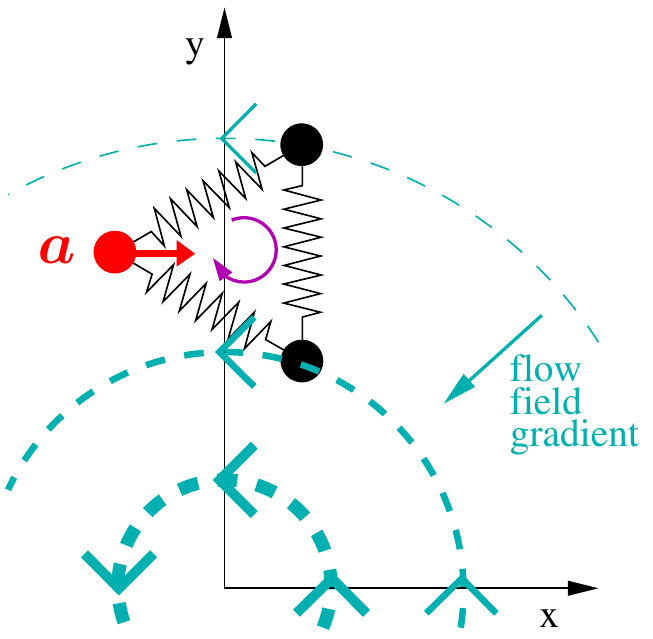}
	\caption{\rev{(Color online)} 
        Illustration of a rotational event in agreement with the strong descent of the curve in Fig.~\ref{fig:angle_3PS_active}. In the depicted state, the active bead tries to ``pin'' the swimmer in its present position: the active drive is directed against the flow field. Since the inner passive bead experiences a higher drag than the outer passive bead due to the radial gradient in the flow velocity, it can be effectively rotated around the ``pinning'' active site. Thus a net rotation of the swimmer with a sense opposite to the rotation of the swirl flow results, as indicated by the curved arrow within the $3$-BS. During the rotation, the outer passive bead is pulled along due to the spring connection to the inner passive bead. 
        }
	\label{fig:pinned-rotation}
\end{figure}

Further analysis reveals that this extreme event to big extent occurs while the active drive is pushing against the swirl flow. The situation is depicted in Fig.~\ref{fig:pinned-rotation}. 
In this state, the active bead tries to ``pin'' the swimmer at its present position when viewed from the laboratory frame. Since the active bead is pushing \rev{from the front}, the situation is unstable. Moreover, the drag acting on the bead closest to the swirl center is higher than for the outer bead due to the radial gradient of the flow velocity. Thus it is conceivable that the swimmer is effectively rotated in this state. The rotational sense of this event is opposite to the rotational sense of the swirl flow, in agreement with the orbital rotations observed in 
Figs.~\ref{fig:rotating_egg} and \ref{fig:rosette}. 

\begin{figure}
	\includegraphics[width=7.5cm]{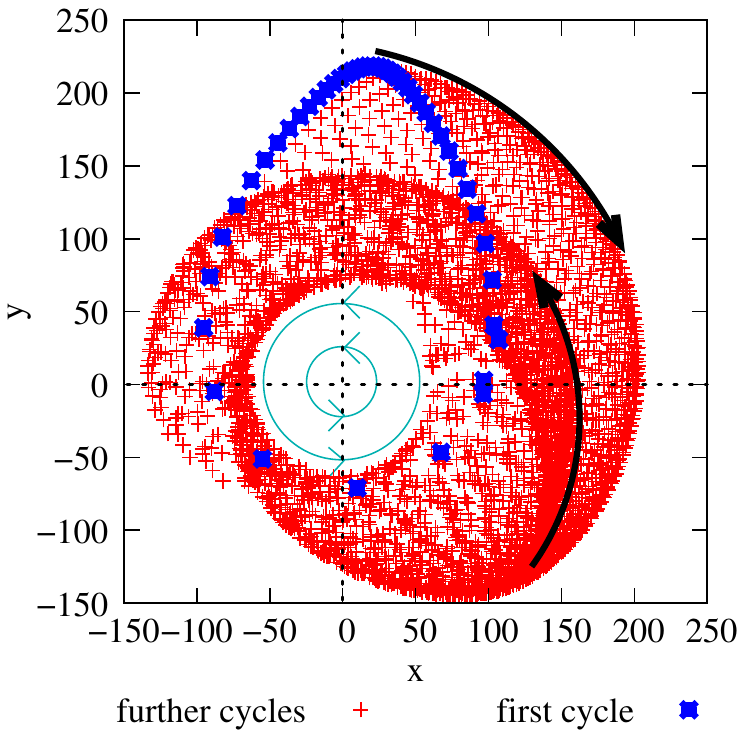}
	\caption{\rev{(Color online)} Center-of-mass trajectory of the active $3$-BS of Figs.~\ref{fig:rotating_egg} and \ref{fig:rosette} in the swirl, observed over a longer time interval. Initially, the egg-shaped orbit rotates with a sense opposite to the rotational sense of the swirl flow. However, with proceeding time, the swimmer is drawn into the swirl towards the swirl center. At a certain distance from the swirl center, the rotational sense of the egg reverses. [Parameters: 
          $a = 0.01$, $k=0.1$.]} 
	\label{fig:volute}
	\end{figure}

Overall, the described mechanism should become less effective, if the swimmer is less effectively ``pinned''. This is the case when the flow field becomes stronger in comparison to the active drive. Over time, as for the passive $3$-BS, the active $3$-BS is drawn into the swirl towards the swirl center. The flow field becomes stronger on this path due to the radial gradient. Indeed, at a certain distance, we observe a reversal in the sense of the net swimmer rotation. As a consequence, also the rotational sense of the egg-shaped orbit reverses, see Fig.~\ref{fig:volute}. 
Then the dominant rotational mechanism and the rotational sense agree with those of the passive $3$-BS. For an active $3$-BS of $a<0$, we did not observe the inverted sense of rotation. In that case, the ``pinning'' bead is \rev{at the rear}, leading to a more stable situation than the one depicted in Fig.~\ref{fig:pinned-rotation}.

\subsection{Active three-bead microswimmer scattered by the swirl}
\label{active-scattered}

\begin{figure}
	\includegraphics[width=8.5cm]{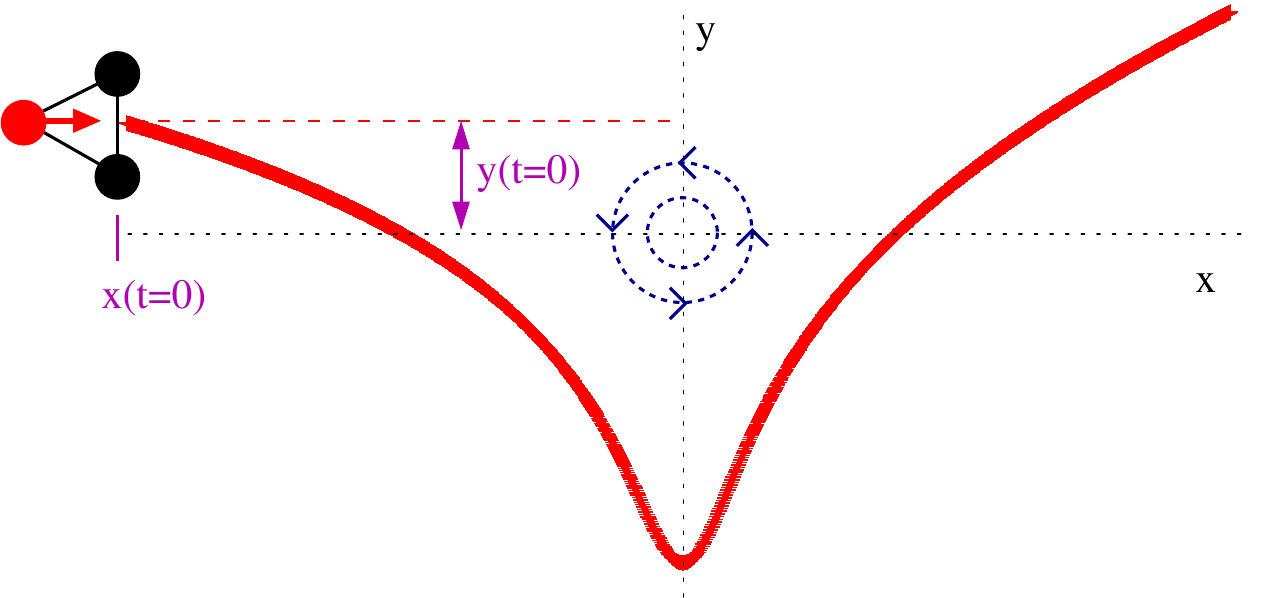}
	\caption{\rev{(Color online)} Set-up and example trajectory of the scattering process of an active $3$-BS in the swirl flow in analogy to a classical scattering experiment. At time $t=0$, the microswimmer is located at coordinates $x_0=x(t=0)$ and $y_0=y(t=0)$ with an active drive parallel to the $x$ axis. 
	[Parameters: \rev{$a = 0.1$}, $k=1$, $x_0= -1000$, $y_0= 20$.]} 
	\label{fig:scatter_set_up}
	\end{figure}

Finally, we address situations in which the active $3$-BS is initially heading towards the 
swirl 
starting from a certain remote distance. Two qualitatively different events may occur. Either the swimmer is captured by the swirl and afterwards drawn towards the swirl center as observed in section \ref{drawn_active_3BS}; thus the swimmer may get drowned. 
Or it is scattered and afterwards can escape from the swirl. 
We depict the corresponding set-up in Fig.~\ref{fig:scatter_set_up}

At time $t=0$, the swimmer starts from an initial position of coordinates $x_0=x(t=0)$ and $y_0=y(t=0)$ with an initial orientation of the active drive parallel to the $x$ axis. We refer to the subsequent event as scattering, if the swimmer reaches a distance from the swirl center \rev{three times} as large as the initial distance without needing more than \rev{twenty cycles} around the swirl to achieve this distance. Otherwise, the swimmer is considered to get captured, \rev{at least transiently}. 
The set-up reminds of a classical Rutherford scattering experiment. We refer to $y_0$ as the impact parameter. 

\begin{figure}
	\includegraphics[width=8.3cm]{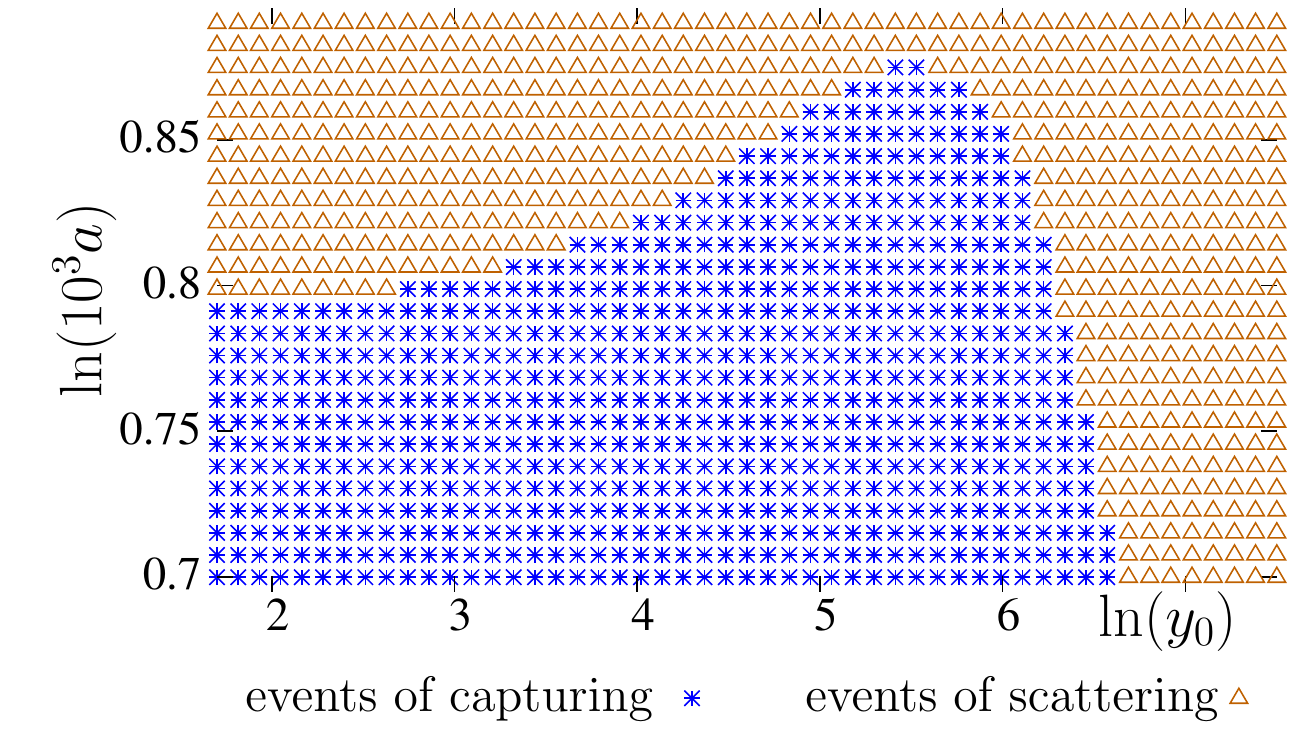}
	\caption{\rev{(Color online)} ``Event diagram'' for the set-up illustrated in Fig.~\ref{fig:scatter_set_up}. We distinguish between capturing and scattering events as a function of the strength of active drive $a$ and the impact parameter $y_0$. Stars (blue) indicate events of getting captured, triangles (brown) mark scattering processes. When the impact parameter $y_0$ is increased at a constant intermediate value of the active drive $a$, we can observe reentrant behavior of the scattering process.
	[Parameters: $x_0=-500$, $k=5$.] 
	}	
	\label{fig:phase_diagram}
	\end{figure}

As in a previous analysis using a different model of a deformable active microswimmer \cite{tarama2014deformable} 
we first 
only vary the strength of the active drive $a$ as well as the 
impact parameter $y_0$, see Fig.~\ref{fig:phase_diagram}.
The initial abscissa $x_0$ and the deformability of the swimmer, quantified by the spring constant $k$, are kept constant. As a result and as it appears plausible, a stronger active drive generally helps the swimmer to escape from the swirl, see Fig.~\ref{fig:phase_diagram}. Capturing is most efficient when the swimmer starts from slightly positive values of the impact parameter $y_0$. The pronounced bump in Fig.~\ref{fig:phase_diagram} indicates an optimal impact parameter for getting captured around $\ln(y_0)\approx5.5$. This shift to positive values of $y_0$ follows from the long-ranged influence of the swirl flow that bends the trajectory downwards in our set-up, see Fig.~\ref{fig:scatter_set_up}. In Fig.~\ref{fig:phase_diagram}, along a horizontal line of constant active drive $a$, we thus find reentrance of the scattering events.  

Finally, we place the active $3$-BS closer to the swirl center and test whether the initial positions lead to capturing or scattering, i.e.\ whether the swimmer gets drowned by the swirl or can escape. 
The strength of the active drive $a$ and the deformability of the swimmer, quantified by the spring constant $k$, are kept constant. 
	\begin{figure}
	\includegraphics[width=6.3cm]{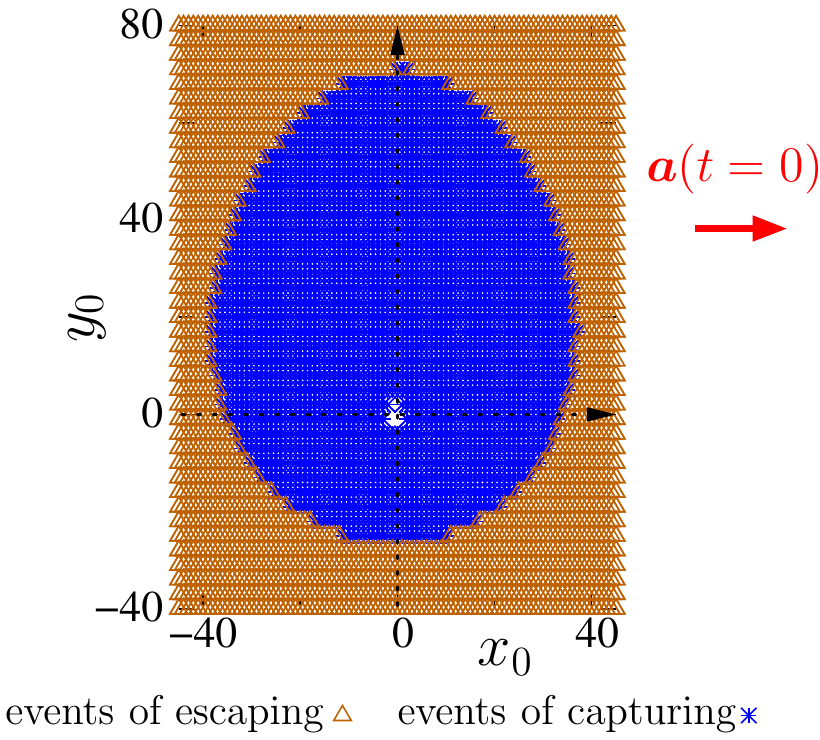}
	\caption{\rev{(Color online)} Capturing and escape events resulting from different initial locations $x_0$ and $y_0$ of the active $3$-BS. Stars (blue) indicate events of getting captured, triangles (brown) indicate escape processes. The non-spherical shape of the area of initial positions leading to capturing events results from the initial orientation of the active drive $\boldsymbol{a}(t=0)$ as indicated on the right. 
	[Parameters: \rev{$a = 0.03$}, $k = 5$.] 
    }
	\label{fig:separatrix}
	\end{figure}
Fig.~\ref{fig:separatrix} confirms that initial locations closer to the swirl center lead to capturing while more remote initial positions allow an escape. The non-circular shape of the area of capturing events highlights the role of the initial orientation of the active drive. 

\section{Deformable four-bead microswimmer in the swirl}\label{fourbead}

Our approach introduces a simplified procedure to construct and investigate well-defined finitely sized model microswimmers in external flow fields. The $2$-BS is special due to its linear extension. It senses the gradient of the external flow field in only one spatial direction. In the swirl flow, it can avoid the gradient by aligning with the flow lines. From this point of view, we expect the results for the two-dimensionally extended $3$-BS to be more generic 
than those of the $2$-BS. To test this conjecture, we briefly investigated the behavior of the $4$-BS introduced in Fig.~\ref{fig:swimmer_model}. 

\begin{figure}
\includegraphics[width=8.3cm]{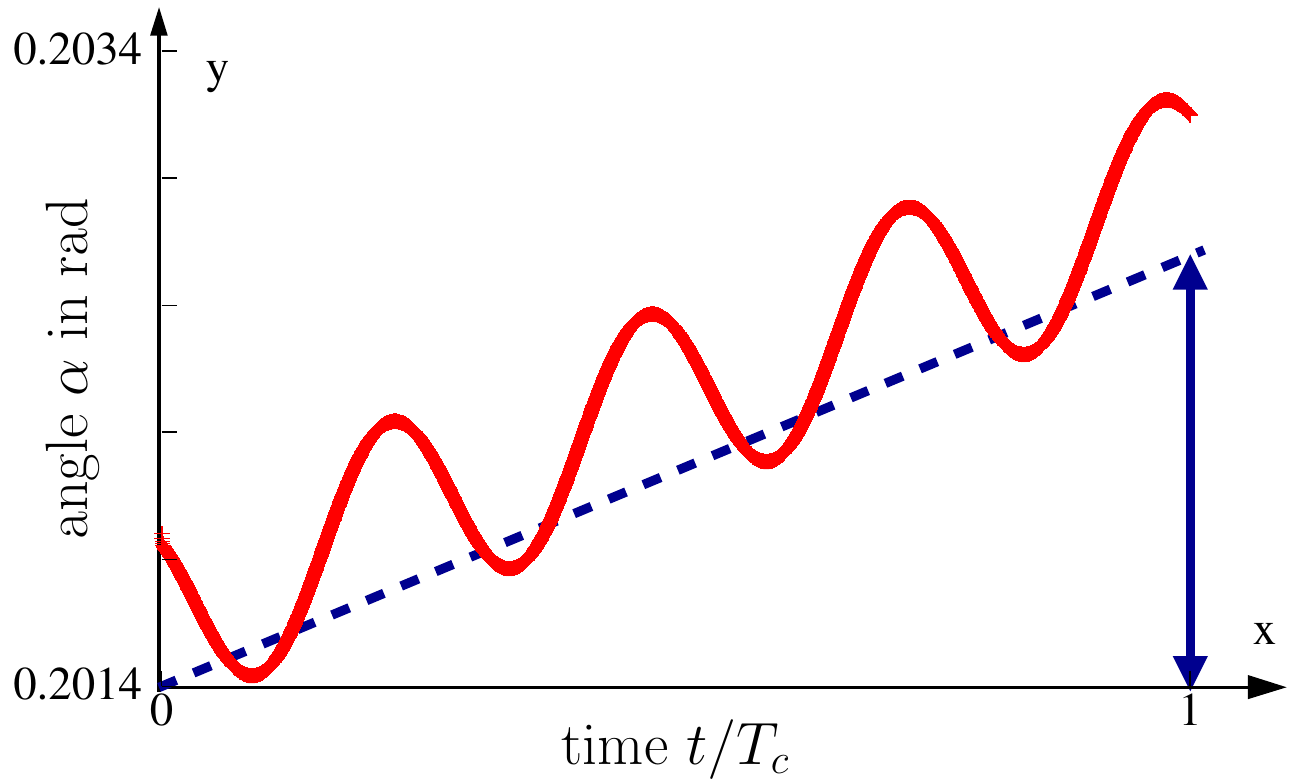}
\caption{\rev{(Color online)} Angular orientation $\alpha$ of a passive $4$-BS as a function of time $t$. Again, $T_c$ denotes the time necessary for the swimmer to circle around the swirl once. In agreement with the swimmer symmetry, we now observe four tiny sinusoidal oscillations in the angular orientation during each cycle. As for the passive $3$-BS, a net rotational drift takes place over time, marked by the dashed line. [Parameters: 
        $a=0$, $k = 0.1$, $R(t=0)=22$, $T_c \approx 3040$.]}
\label{fig:winkel_osz4PS}
\end{figure}

In analogy to Figs.~\ref{fig:angle_3PS} and \ref{fig:dreieck_symmetrie}, we expect for the passive $4$-BS a small net rotational drift in the swirl flow. Moreover, we anticipate tiny four-fold orientational oscillations during each cycle around the swirl center, due to the four-fold rotational symmetry of the passive $4$-BS. Indeed, Fig.~\ref{fig:winkel_osz4PS} confirms our expectations.
Apart from that, as for the passive $3$-BS, we find that the passive $4$-BS is drawn into the swirl. 

\begin{figure}
	\includegraphics[width=7.3cm]{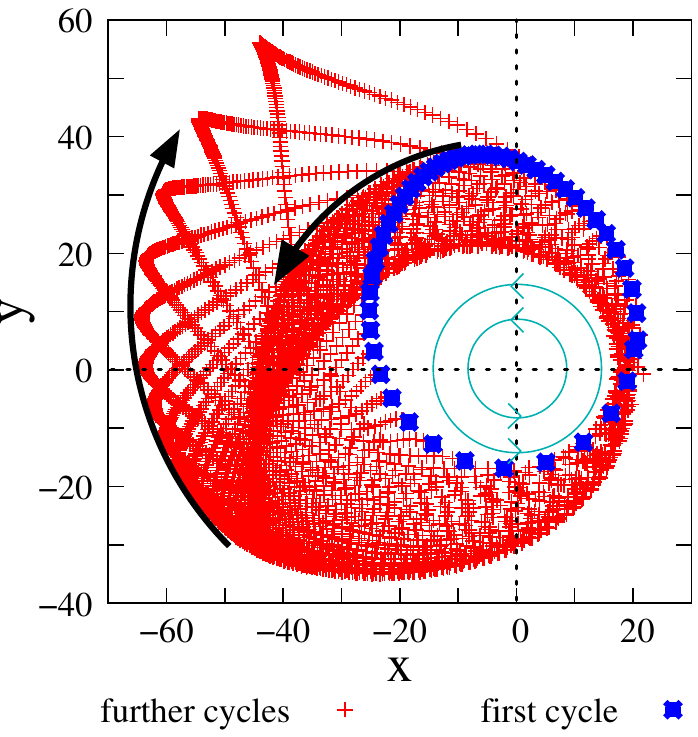}
	\caption{\rev{(Color online)} Center-of-mass trajectory of an active $4$-BS. 
          As for the active $3$-BS, the orbit around the swirl center is distorted to an egg-like shape. Over time this egg rotates. In the present case, the strength of the active drive $a$ is high enough so that the swimmer drifts outwards, away from the swirl center. During this process, the rotational sense of the egg reverses. 
          [Parameters: 
          $a = 0.056$, $k=0.1$.]
          } 
	\label{fig:drehungs_aenderung4BS}	
\end{figure}

Considering an active $4$-BS, 
we observe that the orbit is distorted to an egg-like shape, similarly to the $3$-BS. An example is shown in Fig.~\ref{fig:drehungs_aenderung4BS}. 
Moreover, as indicated in the figure, we likewise find rotations of this egg-shaped orbit. In the depicted case, the strength of the active drive $a$ is high enough so that the swimmer over time increases its distance from the swirl center. During this process we can observe an inversion of the sense of rotation of the egg-shaped orbit, similarly to the case of the $3$-BS in Fig.~\ref{fig:volute}. However, in our example of the $4$-BS in Fig.~\ref{fig:drehungs_aenderung4BS}, the drift of the swimmer is outwards instead of inwards. We observe in Fig.~\ref{fig:drehungs_aenderung4BS} a change of rotational sense oppositely to the one in Fig.~\ref{fig:volute}. 

It is our understanding that the observed peculiarities, particularly the drag towards the swirl center and the net orientational drifts, result from the non-vanishing extension of our regular bead-spring microswimmers. 
Thus we expect that other regular bead-spring swimmers composed of a higher number of beads ($M>4$) show analogous behavior as found for the $3$-BS and the $4$-BS. Yet, this hypothesis and the detailed influence of additional beads and springs will have to be clarified in the future.

\section{Discussion and possible extensions of the model}\label{flucthydro}


As a benefit of our simplified paradigmatic approach, effects resulting from the finite extension and deformability of the swimmers are readily isolated and identified. 
For instance, due to the finite spatial extension, the gradients in the surrounding external flow field are directly sensed by the swimmers with important consequences. One example is that $M$-BSs with $M=3,4$ and weak active drive are drawn into the swirl. This effect has not been found in a previous model \cite{tarama2014deformable}, where the finite spatial extension of the swimmer has not been taken into account. 

In the following, we list some possible extensions of our minimal model. 
It is straightforward to include additional features of realistic systems, which in more elaborate models often requires high computational effort. 
For instance, one could take into account thermal fluctuations of the beads in the present model. Corresponding stochastic noise terms then must be added to Eq.~(\ref{eq_motion_unscaled}). We performed some first tests and naturally found that the trajectories get noisy, but, as long as the fluctuations are not too strong, qualitatively the same effects as described above can be observed. It should be mentioned, however, that also fluctuations of the external flow field should then be taken into account, which requires further knowledge of the specific set-up. 

Moreover, hydrodynamic interactions between the beads of the swimmer could be included. These depend on the boundary conditions under consideration. For passive swimmers, within the bulk of a fluid, and for large enough spatial separation of the beads with respect to their size, hydrodynamic interactions are readily described by Oseen tensors to lowest order \cite{dhont1996introduction}. We have performed according tests and found that hydrodynamic interactions increase on average the degree of deformation in the flow gradient. Close to surfaces, the description of hydrodynamic interactions changes \cite{lauga2009hydrodynamics,spagnolie2012hydrodynamics}. 

\rev{When hydrodynamic interactions are included in the active case, a description in terms of an effective active drive \cite{kummel2014kummel,hagen2015can} of one of the beads is no longer consistent. A microswimmer typically sets the surrounding fluid into motion while achieving self-propulsion \cite{lauga2009hydrodynamics,menzel2015tuned,elgeti2015physics}. These self-induced fluid flows constitute another source of hydrodynamic interaction. Fluid is usually pushed outwards, i.e.\ away from the microswimmer, along certain directions, and pulled inwards, i.e.\ drawn towards the microswimmer, along different directions. If one of the passive beads is located within the window of activated flow away from the active bead, the swimmer is likely to be on average extended along this direction \cite{babel2015dynamics}; in the opposite case, where a passive bead is located within the inward flow, the swimmer is likely to be contracted along that direction \cite{babel2015dynamics}. Importantly, the change in the swimmer extension modifies the ability of the microswimmer to sense gradients in the externally imposed flow field. As one illustrative example, the tendency of the active $2$-BS to spiral outwards in the swirl flow should increase when the swimmer is extended due to hydrodynamic interactions; in contrast to that, it should decrease, when the $2$-BS is on average contracted. It will be interesting to study in the future in more detail the influence of hydrodynamic interactions in this context. For that purpose, as a minimal approach, the swimming mechanism should be modeled using an active force dipole as described e.g.\ in Refs.~\cite{saintillan2012kinetic,jayaraman2012,yeomans2014introduction,babel2015dynamics}.}

So far, we have only investigated the two-dimensional motion of our model microswimmers, and only planar swimmer geometries were considered. Naturally, our concept is easily extended to propulsion in three spatial dimensions, including the effect of imposed three-dimensional flow fields. Tetrahedral model microswimmers would represent the simplest three-dimensional bead-spring swimmer geometries. Moreover, the behavior of less regular swimmer structures could be analyzed, with different bead sizes and thus different hydrodynamic friction constants, varying spring stiffnesses, and varying undeformed spring lengths, or different strengths of self-propulsion of individual beads on the same microswimmer.

\section{Conclusions}\label{conclusions}

In summary, we introduced simplified bead-spring model microswimmers and investigated their behavior in a circular swirl flow. The colloidal beads are connected by harmonic springs, and in the active case one of the beads experiences an active drive. A linear swimmer consisting of two beads aligns with the flow lines and is circularly convected around the swirl. An active drive slowly transports it radially outwards. For a three-bead microswimmer, a small rotational drift together with tiny orientational oscillations are observed while being convected around the swirl. The circular orbit around the swirl is deformed to an egg-like shape when an active drive is switched on. Small net rotations of this egg-shaped orbit are observed, leading to rosette-like trajectories. For low enough magnitude of the active drive, the swimmer is drawn towards the swirl center. Remarkably, the sense of rotation of the egg-shaped orbit can reverse during this process. For higher magnitudes of active drive, we distinguish between capturing and scattering events by the swirl, if the swimmer is initially heading towards the swirl. Aspects of this behavior persist for a swimmer consisting of four beads, with certain variations. This opens the space for further investigations in the future on deformable multi-bead microswimmers. 

It should be possible to analyze the behavior of the proposed types of microswimmers experimentally. 
Colloidal particles can be effectively connected to each other using DNA fragments as linkers \cite{dreyfus2005microscopic}. An active drive can be imposed when an appropriate self-propelling Janus particle \cite{paxton2004catalytic,howse2007self,walther2008janus, jiang2010active,volpe2011microswimmers, buttinoni2012active,theurkauff2012dynamic,buttinoni2013dynamical} is inserted instead of one of the other passive colloidal particles. Different geometries in the form of the investigated two-, three-, and four-bead swimmers could be realized in this way. 
In the simplest case, a two-dimensional motion results when the swimmer is confined to the two-dimensional surface of a fluid. 
We chose the circular swirl flow as an externally imposed fluid flow because it should be easily implemented experimentally. In principle, a cylindrical cavity with a steadily rotating magnetic stir bar at the bottom can be used in a first attempt. 
Our approach may serve as a simplified paradigmatic model to describe the behavior of more complex deformable self-propelled microswimmers in external flow fields. Examples are self-propelled droplets \cite{nagai2005mode,chen2009self,takabatake2011spontaneous,kitahata2011spontaneous, thutupalli2011swarming,yoshinaga2012drift,schmitt2013swimming,yoshinaga2014spontaneous}. This possible connection should be further probed and tested in the future.

On the theoretical side, the simple model microswimmer shall be a starting point for our future investigations of the collective behavior of many interacting deformable swimmers. For this purpose, additional steric inter-swimmer interactions must be imposed to avoid unphysical overlaps. 
In such a situation, our model will show its strength: all interactions, hydrodynamic and steric, together with thermal fluctuations and finite extensions of the individual swimmers can be formulated consistently. 
Simultaneously, the model is simple enough to handle a collection of many interacting particles. This shall inspire direct particle-based computer simulations or the derivation of corresponding statistical theories. 

A variant of our model also currently under investigation are ``active colloidal polymers'' \cite{kaiser2014unusual,harder2014activity,kaiser2015does, shin2015facilitation,chen2015transport}. In this case, many colloidal particles are linked to a linear chain. Two cases are analyzed in this context. On the one hand, the chain can be made of passive colloidal particles placed into a background of self-propelling microswimmers \cite{kaiser2014unusual,harder2014activity,shin2015facilitation}. On the other hand, the chain itself could be composed of self-propelling Janus particles \cite{kaiser2015does,chen2015transport}. Naturally, also combinations of the two cases, or chains that are only partially active, can be studied in analogy to our simplified finitely-sized model microswimmers. 


\vspace{.6cm}

\begin{acknowledgments}
The authors thank the Deutsche Forschungsgemeinschaft (DFG) for support of this work through the priority program SPP 1726 ``Microswimmers''. 
\end{acknowledgments}


\end{document}